\documentclass[aps,prd,preprintnumbers,superscriptaddress,nofootinbib,notitlepage]{revtex4-1}
\usepackage[dvipdfmx]{graphicx}
\usepackage{bm,latexsym,amsmath,amssymb,amsfonts,mathrsfs}
\usepackage{color}
\input{colordvi.tex}
\allowdisplaybreaks[1]
\newcommand*{\D}{\textrm{d}}

\definecolor{pn}{rgb}{1.0,0.2,0.8}

\begin{document}
\title{Effective scalar-tensor description of regularized Lovelock gravity in four dimensions}

\author{Tsutomu~Kobayashi}
\email[Email: ]{tsutomu"at"rikkyo.ac.jp}
\affiliation{Department of Physics, Rikkyo University, Toshima, Tokyo 171-8501, Japan
}
\begin{abstract}
We reformulate the recently proposed regularized version of
Lovelock gravity in four dimensions as a scalar-tensor theory.
By promoting the warp factor of the internal space to
a scalar degree of freedom by means of Kaluza-Klein reduction,
we show that regularized Lovelock gravity
can be described effectively by a certain subclass of the Horndeski theory.
Cosmological aspects of this particular scalar-tensor theory are studied.
It is found that the background with a scalar charge is
generically allowed. The consequences of this scalar charge are
briefly discussed.
\end{abstract}
\pacs{%
04.50.Kd  
}
\preprint{RUP-20-11}
\maketitle
\section{Introduction}

Lovelock gravity~\cite{Lovelock:1971yv} is
the most general metric theory of gravity
in higher dimensions retaining the second-order nature of
field equations for the metric.
The action for
Lovelock gravity in $D$ dimensions is given by
\begin{align}
S=\int\D^{D}x\sqrt{-g_{D}}\sum_{p=0}^{\lfloor (D-1)/2\rfloor}\alpha_p{\cal L}_p,
\end{align}
where $\alpha_p$ is the coupling constant and
\begin{align}
{\cal L}_p:=\frac{1}{2^p}
\delta^{A_1B_1\cdots A_pB_p}_{C_1D_1\cdots C_pD_p}
{\cal R}_{A_1B_1}^{~~~~~C_1D_1}
\cdots
{\cal R}_{A_pB_p}^{~~~~~C_pD_p}.
\end{align}
The first three terms are written more explicitly as
\begin{align}
{\cal L}_0=1,\quad {\cal L}_1={\cal R},\quad
{\cal L}_2={\cal R}^2-4{\cal R}^{AB}{\cal R}_{AB}
+{\cal R}^{ABCD}{\cal R}_{ABCD}.
\end{align}
In $D=4$ dimensions, the Lovelock Lagrangian uniquely reduces to
the Einstein-Hilbert term (${\cal L}_1$) plus a cosmological constant
(${\cal L}_0$). This puts strong limitations on
constructing the metric theory of gravity other than
Einstein in four dimensions.

Recently, a trick has been proposed to circumvent this
limitation~\cite{Glavan:2019inb,Casalino:2020kbt}.
(See also~\cite{Tomozawa:2011gp,Cognola:2013fva} for earlier works.)
The trick amounts to
rescaling the coupling constants $\alpha_p$ ($p\ge 2$)
as
\begin{align}
\alpha_p=\frac{\alpha_p'}{D-4}\label{resalpha1}
\end{align}
and then taking the $D\to 4$ limit.
This procedure leaves nonvanishing contributions in
the gravitational field equations, and thus one ends up with
a seemingly novel theory of gravity in four dimensions.
Although how this ``regularization'' works is not so evident
at the level of the action or the covariant field equations,
an explicit analysis of cosmological solutions,
black hole solutions, and
perturbations~\cite{Glavan:2019inb,Casalino:2020kbt,Fernandes:2020rpa,%
Konoplya:2020qqh,Wei:2020ght,Kumar:2020owy,Doneva:2020ped,Ghosh:2020vpc}
shows that the Lovelock terms yield the factor of $(D-4)$
in the field equations
to cancel $(D-4)$ in the denominator.
See also Refs.~\cite{Konoplya:2020bxa,Guo:2020zmf,%
Hegde:2020xlv,Zhang:2020qew,Singh:2020xju} for aspects of
black hole solutions in this regularized version of Lovelock gravity.

Let us take a look at the cosmological spacetime
studied in~\cite{Glavan:2019inb,Casalino:2020kbt}.
The $D$-dimensional cosmological
metric is assumed to take the form
\begin{align}
g_{AB}\D x^A\D x^B=-\D t^2+a^2(t)
\left(\delta_{ij}\D x^i\D x^j+\delta_{ab}\D x^a\D x^b\right),
\end{align}
where $i,j=1,2,3$ and indices $a,b$ run through
$4, 5, \cdots, 3+n$ with $n=D-4$.
Substituting this metric to the gravitational field equations
and then taking the $D\to 4$ limit along with
rescaling the coupling constants as Eq.~\eqref{resalpha1},
one obtains the modified background equations in four
dimensions~\cite{Glavan:2019inb,Casalino:2020kbt}.
Here some questions arise.
What happens if one assumes a different scale factor
of the $n$-dimensional internal space,
$b^2(t)\delta_{ab}\D x^a\D x^b$?
And then, what is the role of the $(a,b)$
(i.e., the internal-space)
components of
the field equations in the $D\to 4$ limit?
More generically, what is the explicit form of covariant equations
to determine the four-dimensional metric after taking the $D\to 4$ limit?
Can the $D\to 4$ limit be taken consistently
for any metric?
Concerning these points, some criticisms
on the validity of taking the $D\to 4$ limit
have been raised~\cite{Ai:2020peo,Gurses:2020ofy,Mahapatra:2020rds,Tian:2020nzb,Arrechea:2020evj}.

To address these issues, in this short paper
we study the dynamics of regularized Lovelock gravity
by assuming a general four-dimensional metric and
a general warp factor of the internal space.
We show that taking the $D\to 4$ limit after the Kaluza-Klein reduction
leaves a dynamical scalar degree of freedom in four dimensions,
yielding a particular class of scalar-tensor theories
within the Horndeski family.
Although the validity of the original formulation is questioned,
regularized Lovelock gravity can thus be reformulated
in a consistent way as a well-defined four-dimensional theory.
Employing this scalar-tensor reformulation,
we revisit the dynamics of a cosmological spacetime
in regularized Einstein-Gauss-Bonnet gravity in four dimensions.

The rest of the paper is organized as follows.
In the next section we clarify the relation between
the Horndeski theory and the $D\to 4$ limit of the Lovelock theory
to provide the scalar-tensor reformulation of the latter theory.
We then study the background dynamics and linear perturbations
of a cosmological spacetime based on the scalar-tensor reformulation
in Sec.~III. A brief summary of the paper is presented in Sec.~IV.

\section{Horndeski and regularized Lovelock}

For simplicity,
we focus on the case of Einstein-Gauss-Bonnet gravity for the moment.
Let us consider the $(4+n)$-dimensional metric of the form
\begin{align}
g_{AB}\D x^A\D x^B=
g_{\mu\nu}(x)\D x^\mu\D x^\nu +e^{2\chi(x)}\D \sigma_K^2,\label{4nmet}
\end{align}
where
$\mu,\nu=0,1,2,3$ and
$\D \sigma_K^2$ is the line element of a
$n$-dimensional maximally symmetric space with constant curvature $K$.
We perform a Kaluza-Klein reduction
starting from the metric~\eqref{4nmet}.
Substituting Eq.~\eqref{4nmet} to the Gauss-Bonnet term and
doing integration by parts, we obtain~\cite{VanAcoleyen:2011mj,Charmousis:2012dw}
\begin{align}
\sqrt{-g_{n+4}}{\cal L}_2
=n\sqrt{-g}\left(
-6K^2e^{-4\chi}+24Ke^{-2\chi}X-2Ke^{-2\chi}R
+8X^2+8X\Box\chi +4G^{\mu\nu}\chi_\mu\chi_\nu+\chi{\cal G}
\right)+{\cal O}(n^2),
\end{align}
where $\chi_\mu=\nabla_\mu\chi$, $X=-\chi_\mu\chi^\mu/2$,
and ${\cal O}(n^2)$ stands for the terms proportional to
$n^2$ (or higher powers of $n$).
Here, $R$ and $G_{\mu\nu}$ are the four-dimensional Ricci
scalar and Einstein tensor, respectively, and
${\cal G}$ is the Gauss-Bonnet combination of the
four-dimensional curvature tensors,
${\cal G}=R^2-4R_{\mu\nu}R^{\mu\nu}+R_{\mu\nu\rho\sigma}R^{\mu\nu\rho\sigma}$.
By rescaling the coupling constant $\alpha_2$ as $\alpha_2=\alpha_2'/n$
and taking the $n\to 0$ limit, we arrive at the
following alternative description of
regularized Gauss-Bonnet gravity,
\begin{align}
{\cal L}=\alpha_0+\left(\alpha_1-2\alpha_2'Ke^{-2\chi} \right)R
 +\alpha_2'
\left[
-6K^2e^{-4\chi}+24Ke^{-2\chi}X+8X^2
+8X\Box\chi +4G^{\mu\nu}\chi_\mu\chi_\nu
+\chi{\cal G}
\right].\label{GB4dH}
\end{align}
As long as the $(4+n)$-dimensional metric in underlying
Einstein-Gauss-Bonnet gravity is assumed to be of the form of Eq.~\eqref{4nmet},
the $n\to0$ ($D\to 4$) limit can be taken consistently
and straightforwardly
for any four-dimensional metric $g_{\mu\nu}$,
giving the above effective Lagrangian.\footnote{After the first version of
this paper was submitted, it was pointed out that the same scalar-tensor theory
(with $K=0$)
is obtained also by introducing a counter term to the action
to cancel the divergence in the $D\to 4$ limit~\cite{Fernandes:2020nbq,Hennigar:2020lsl}.
Our result is also consistent with the analysis of amplitudes~\cite{Bonifacio:2020vbk}.}
This is one of the main result of this paper.
In the case of $K=0$, the theory has the invariance
under $\chi\to\chi+\,$const.

This theory can be viewed as a particular subclass of
the Horndeski theory~\cite{Horndeski:1974wa} (see~\cite{Kobayashi:2019hrl}
for a review).
The Lagrangian of the Horndeski theory is the sum of the
following four Lagrangians~\cite{Deffayet:2011gz,Kobayashi:2011nu},
\begin{align}
{\cal L}_2^H\{{G_2}\}&:=G_2(\chi,X),\\
{\cal L}_3^H\{G_3\}&:=-G_3(\chi,X)\Box\chi,\\
{\cal L}_4^H\{G_4\}&:=G_4(\chi,X)R+G_{4,X}\delta^{\mu_1\mu_2}_{\nu_1\nu_2}
\chi_{\mu_1}^{\nu_1}\chi_{\mu_2}^{\nu_2},\\
{\cal L}_5^H\{G_5\}&:=G_5(\chi,X)G^{\mu\nu}\chi_{\mu\nu}
-\frac{1}{6}G_{5,X}\delta^{\mu_1\mu_2\mu_3}_{\nu_1\nu_2\nu_3}
\chi_{\mu_1}^{\nu_1}\chi_{\mu_2}^{\nu_2}\chi_{\mu_3}^{\nu_3}.
\end{align}
The Lagrangian~\eqref{GB4dH} corresponds to the case with
\begin{align}
G_2&=\alpha_0+\alpha_2'\left(-6K^2e^{-4\chi}+24Ke^{-2\chi}X+8X^2\right),
\\
G_3&=-8\alpha_2'X,
\\
G_4&=\alpha_1+\alpha_2'\left(-2Ke^{-2\chi}+4X\right),
\\
G_5&=-4\alpha_2'\ln X.
\end{align}
Here, we used the fact that
$G^{\mu\nu}\chi_\mu\chi_\nu$ can be written equivalently as
$XR+\delta^{\mu_1\mu_2}_{\nu_1\nu_2} %
\chi_{\mu_1}^{\nu_1}\chi_{\mu_2}^{\nu_2}$ up to total derivatives.
Note also that the nonminimal coupling to the Gauss-Bonnet term can be
reproduced from the Horndeski functions including the $\ln X$
terms~\cite{Kobayashi:2011nu}.

The gravitational field equations derived from the Lagrangian~\eqref{GB4dH}
take the form
\begin{align}
-\frac{\alpha_0}{2}\delta_\mu^\nu + \alpha_1 G_\mu^\nu +\alpha_2' H_\mu^\nu=\frac{1}{2}
T_\mu^\nu,\label{eomgmn}
\end{align}
where $H_\mu^\nu$ is a $\chi$-dependent tensor and
$T_{\mu}^\nu$ is the energy-momentum tensor of matter.
One of the interesting properties of the above scalar-tensor theory
is that a particular linear combination of
the $\chi$-field equation of motion
and the trace of Eq.~\eqref{eomgmn} reduce to
\begin{align}
-4\alpha_0-2\alpha_1R-\alpha_2'{\cal G}=T_\mu^\mu.
\label{tracegb}
\end{align}
One can thus remove $\chi$ from the trace part of the
gravitational field equations.
Equation~\eqref{tracegb} itself is as expected from the
original formulation of the theory~\cite{Glavan:2019inb}, and
our scalar-tensor reformulation can correctly reproduce
the same result, though this equation is not trivial
from the scalar-tensor viewpoint.

Rewriting explicitly the $p\ge 3$ Lovelock terms as the Horndeski theory
is much more involved, though they must reduce to the form of
the second-order scalar-tensor theory anyway~\cite{VanAcoleyen:2011mj}.
For example, substituting the metric~\eqref{4nmet}
with $K=0$
to ${\cal L}_3$ yields
\begin{align}
\frac{{\cal L}_3}{n}&=
-6X{\cal G}-6 \delta^{\mu_1\cdots\mu_4}_{\nu_1\cdots\nu_4}
R_{\mu_1\mu_2}^{~~~~~\nu_1\nu_2}\chi_{\mu_3}^{\nu_3}\chi_{\mu_4}^{\nu_4}
\notag \\ &\quad
-12\delta^{\mu_1\mu_2\lambda_1\lambda_2}_{\lambda_3\lambda_4\mu_3\mu_4}
R_{\lambda_1\lambda_2}^{~~~~~\lambda_3\lambda_4}\chi_{\mu_1}\chi^{\mu_3}
\chi_{\mu_2}^{\mu_4}
\notag \\ & \quad
+192X^3-288X\nabla_\mu X\chi^\mu
+{\cal L}_4^H\{-72X^2\}+{\cal L}_5^H\{96X\}
\notag \\ & \quad
+96X\left(G^{\mu\nu}\chi_\mu\chi_\nu+\delta^{\mu_1\mu_2}_{\nu_1\nu_2}
\chi_{\mu_1}^{\nu_1}\chi_{\mu_2}^{\nu_2}\right)-96\Box\chi \nabla_\mu X\chi^\mu
-96\nabla_\mu X\nabla^\mu X
+{\cal O}(n).
\end{align}
The first line is of the form of the generalized Galileon~\cite{Deffayet:2011gz},
\begin{align}
{\cal L}_{6}^{\rm Gal}
=\delta_{\nu_1\cdots\nu_4}^{\mu_1\cdots\mu_4}\left[
\frac{3}{4}G_6(\chi,X)R_{\mu_1\mu_2}^{~~~~~\nu_1\nu_2}R_{\mu_3\mu_4}^{~~~~~\nu_3\nu_4}
+3G_{6,X}R_{\mu_1\mu_2}^{~~~~~\nu_1\nu_2}\chi_{\mu_3}^{\nu_3}\chi_{\mu_4}^{\nu_4}
+G_{6,XX}\chi_{\mu_1}^{\nu_1}\cdots \chi_{\mu_4}^{\nu_4}
\right],
\end{align}
which is a total derivative in four dimensions.
The second line, the derivative coupling to the double dual Riemann
tensor, can be written equivalently as ${\cal L}_5\{-48X\}$~\cite{Kobayashi:2013ina}.
It is easy to rearrange the other terms and we finally obtain,
in the $n\to0$ limit,
\begin{align}
\alpha_3{\cal L}_3\to \alpha_3'\left[
{\cal L}_{2}^H\{192X^3\}+{\cal L}_3^{H}\{-144 X^2\}
+{\cal L}_4^H\{24X^2\}+{\cal L}_5^H\{48X\}
\right].
\end{align}
Similarly, the $p$-th Lovelock term yields
${\cal L}_2^H\{X^p\}$, ${\cal L}_3^H\{X^{p-1}\}$, ${\cal L}_5^H\{X^{p-1}\}$,
and ${\cal L}_2^H\{X^{p-2}\}$ with particular coefficients.
This confirms that the contributions from the higher-order Lovelock terms
are high-energy corrections to Eq.~\eqref{GB4dH}.
It is straightforward to include the curvature $K$.

\section{Revisiting Cosmology in regularized Einstein-Gauss-Bonnet Gravity}

Let us study the cosmological dynamics, focusing
again on the case of
regularized Einstein-Gauss-Bonnet gravity
and its scalar-tensor reformulation~\eqref{GB4dH} with $K=0$
for simplicity.
Since $\chi$ is promoted to be a dynamical field
in our scalar-tensor reformulation,
we will emphasize its consequences on the background and
perturbation dynamics.

\subsection{Background Cosmology}

For the flat Friedmann-Lema\^{i}tre-Robertson-Walker metric,
$\D s^2=-\D t^2+a^2(t)\delta_{ij}\D x^i\D x^j$, and $\chi=\chi(t)$,
we have
\begin{align}
H_0^0&=3\dot\chi^4-12H\dot\chi^3+18H^2\dot\chi^2-12H^3\dot\chi,
\label{Heq1}
\\
H_i^j&=\left(-\dot\chi^4+6H^2\dot\chi^2-8H^3\dot\chi
-4\dot\chi^2\ddot\chi +4\dot H\dot\chi^2
+8H\dot\chi\ddot\chi
-4H^2\ddot\chi -8H\dot H\dot\chi
\right)\delta_i^j,\label{Heq2}
\end{align}
where $H=\dot a/a$ is the Hubble parameter and
a dot stands for differentiation with respect to $t$.
The $\chi$-field equation, or, equivalently,
$\alpha_2'\nabla_\nu H_\mu^\nu=0$,
reduces to
\begin{align}
  \frac{\alpha_2'}{a^3}\frac{\D}{\D t}
  \left[a^3(\dot\chi-H)^3\right]=0.
\end{align}
This can be integrated to give
\begin{align}
\dot\chi =H+\frac{{\cal C}}{a},\label{solchi}
\end{align}
where the integration constant ${\cal C}$ is the scalar charge
associated with the shift symmetry $\chi\to\chi+\,$const.
The case with ${\cal C}=0$ corresponds to
the isotropically expanding solution,
$\D s^2=-\D t^2+a^2\delta_{ij}\D x^i\D x^j+a^2\delta_{ab}\D x^a\D x^b$,
which is assumed from the beginning in~\cite{Glavan:2019inb}.
The scalar-tensor reformulation reveals that
the background with the nonvanishing scalar charge is in fact allowed.
In an accelerating universe, the second term in Eq.~\eqref{solchi}
decays quickly compared to the first, so that $\dot \chi = H$ is
a dynamical attractor.
However, this is not the case in a decelerating universe.

Substituting the solution~\eqref{solchi} to Eqs.~\eqref{Heq1} and~\eqref{Heq2},
we now have the following background cosmological equations,
\begin{align}
\alpha_0+6(\alpha_1H^2+\alpha_2'H^4)&=\rho+\frac{6\alpha_2'{\cal C}^4}{a^4},
\\
-4\alpha_1\Gamma \dot H &= \rho+P+\frac{8\alpha_2'{\cal C}^4}{a^4},
\end{align}
where
\begin{align}
\Gamma:=1+\frac{2\alpha_2'H^2}{\alpha_1},
\end{align}
and $\rho$ and $P$ are the energy density and pressure of matter, respectively.
It can be seen that the nonvanishing scalar charge
gives rise to an extra radiation-like component in the background equations.
This is a specific nature of the particular scalar-tensor theory~\eqref{GB4dH}
with $K=0$.
In the case of ${\cal C}=0$
the background equations
derived in~\cite{Glavan:2019inb} are reproduced correctly.

\subsection{Cosmological Perturbations}

Let us move to the perturbation dynamics.\footnote{It is worth emphasizing that
we consider the perturbation dynamics within our scalar-tensor reformulation
of the original theory. This effectively kills some of the perturbation modes
associated to the internal space such as the Kaluza-Klein vector modes.
In the presence of such modes, there is no guarantee that one can take
the $D\to 4$ limit consistently after perturbing the full $D$-dimensional metric.}
As a matter field we simply add a canonical scalar field
described by the Lagrangian ${\cal L}_\phi=-\phi_\mu\phi^\mu/2-V(\phi)$.
We fix the temporal gauge degree of freedom by imposing that
$\phi(t,x^i)=\phi(t)$.
The remaining spatial gauge degrees of freedom can be used to
write the metric as
\begin{align}
\D s^2=-N^2\D t^2+\gamma_{ij}(\D x^i+N^i\D t)(\D x^j+N^j\D t),
\end{align}
where
\begin{align}
N=1+\delta N,\quad N_i=\partial_i\psi,\quad \gamma_{ij}=a^2e^{2\zeta}(e^h)_{ij},
\end{align}
with $(e^h)_{ij}=\delta_{ij}+h_{ij}+h_{ik}h_{kj}/2+\cdots$.
The $\chi$-field also fluctuates,
\begin{align}
\chi=\ln a +{\cal C}\int\frac{\D t}{a}+\delta \chi.
\end{align}
Since all the gauge degrees of freedom have already been fixed,
we cannot gauge away $\delta\chi$.
One would therefore expect that
there are two dynamical modes in the scalar sector.

It is straightforward to compute the quadratic Lagrangian for
the tensor modes $h_{ij}$:
\begin{align}
\sqrt{-g}{\cal L}= 2a^3\left[\left(
\alpha_1\Gamma-\frac{2\alpha_2'{\cal C}^2}{a^2}\right)
\dot h_{ij}^2-\left(\alpha_1\Gamma+4\alpha_2'\dot H+\frac{2\alpha_2'{\cal C}^2}{a^2}\right)
\frac{(\partial h_{ij})^2}{a^2}
\right].\label{tensorquad}
\end{align}
Clearly, this Lagrangian
reproduces the equation of motion presented in~\cite{Glavan:2019inb}
for ${\cal C}=0$.
One might worry about ghost instabilities
for a sufficiently large scalar charge. However,
even if the scalar charge is as large as ${\cal C}/a\gtrsim H$,
we have, from the background equations, that
$\alpha_1\Gamma H^2\sim \alpha_2'{\cal C}^4/a^4\gtrsim \alpha_2'H^2{\cal C}^2/a^2$,
which implies that ghost instabilities in the tensor sector are
generically avoided.

The propagation speed of tensor modes $c_{\rm GW}$
is strongly constrained at low redshifts
by the observation of GW170817 and its electromagnetic
counterparts~\cite{TheLIGOScientific:2017qsa,Monitor:2017mdv}
as $|c_{\rm GW}^2-1|\lesssim 10^{-15}$.
From Eq.~\eqref{tensorquad} we see that the propagation speed
is given by
\begin{align}
c_{\rm GW}^2=\frac{\alpha_1\Gamma+4\alpha_2'\dot H+2\alpha_2'{\cal C}^2/a^2}%
{\alpha_1\Gamma-2\alpha_2'{\cal C}^2/a^2},
\end{align}
which deviates from 1 in general,
$|c_{\rm GW}^2-1|\sim \alpha_2'\dot H/\alpha_1, \alpha_2'{\cal C}^2/\alpha_1a^2
(\lesssim (\alpha_2')^{1/2}H/\alpha_1^{1/2})$. However, since
$H/\alpha_1^{1/2}$ is as small as $10^{-60}$ in the present Universe (where we
substituted the Planck mass to $\alpha_1^{1/2}$), this does not
lead to a meaningful constraint on the dimensionless parameter $\alpha_2'$.

The quadratic Lagrangian for the scalar perturbations reads
\begin{align}
\sqrt{-g}{\cal L}&=a^3\Biggl[
\left(\frac{\dot\phi^2}{2}-6\alpha_1\Gamma H^2
+\frac{12\alpha_2'{\cal C}^4}{a^4}\right)\delta N^2
-\frac{2AH}{a^2}\delta N \partial^2\psi
+\frac{2B}{a^2}\dot\zeta\partial^2\psi
+6AH \delta N\dot\zeta
\notag \\ &\quad
-\frac{2B}{a^2}
\delta N\partial^2\zeta
-3B\dot\zeta^2+\frac{2}{a^2}
\left(\alpha_1\Gamma + 4\alpha_2\dot H+\frac{2\alpha_2'{\cal C}^2}{a^2}\right)
(\partial\zeta)^2
\notag \\ & \quad
+\frac{4\alpha_2'{\cal C}^2}{a^2}\left(
\frac{2}{a^2}\dot{\delta\chi}\partial^2\psi
+3\dot{\delta\chi}^2-\frac{1}{a^2}(\partial\delta\chi)^2
-\frac{2}{a^2}\delta N\partial^2\delta\chi-6\dot{\delta\chi}\dot\zeta
\right)
\notag \\ & \quad
+\frac{8\alpha_2'{\cal C}^3}{a^3}\left(
\frac{1}{a^2}\delta\chi\partial^2\psi-3\delta N\dot{\delta\chi}
+3\zeta\dot{\delta\chi}
\right)
\Biggr],\label{scalar2ac}
\end{align}
where
\begin{align}
A&:=2\alpha_1\Gamma +\frac{4\alpha_2'{\cal C}^3}{a^3H},
\\
B&:=2\alpha_1\Gamma-\frac{4\alpha_2'{\cal C}^2}{a^2}.
\end{align}
The variation with respect to $\psi$ gives
\begin{align}
\delta N=\frac{B}{A}\frac{\dot\zeta}{H}
+\frac{4\alpha_2'{\cal C}^2}{A a^2}\frac{\dot{\delta\chi}}{H}
+\frac{4\alpha_2'{\cal C}^3}{A a^3H}\delta\chi.
\end{align}
Substituting this back to the above Lagrangian, we obtain
the quadratic Lagrangian written in terms of $\zeta$ and $\delta\chi$.
The general expression is messy, and hence we expand the Lagrangian
in terms of ${\cal C}$ assuming that the
contribution of the scalar charge to the background is subdominant,
leading to
\begin{align}
\sqrt{-g}{\cal L}\simeq 2\alpha_1 a^3 \Gamma \epsilon  \left[\dot\zeta^2
-\frac{(\partial\zeta)^2}{a^2}\right]+
12\alpha_2'{\cal C}^2a\left[
\dot{\delta\chi}^2-\frac{(\partial\delta\chi)^2}{3a^2}
\right],
\end{align}
where $\epsilon:=-\dot H/H^2$.
One sees that the sound speed of $\delta\chi$
is given by $1/\sqrt{3}$, signaling its radiation-like nature.
On the ${\cal C}=0$ background, $\delta\chi$ apparently
drops off from the Lagrangian
and thereby
the equation of motion in~\cite{Glavan:2019inb} is reproduced
from our Lagrangian.
Note, however, that the ${\cal C}\to 0$ limit must be taken with care
because the disappearance of the kinetic term of $\delta\chi$
would indicate a strong coupling.\footnote{The analysis of amplitudes
shows that the scalar degree of freedom is strongly coupled around a
flat space~\cite{Bonifacio:2020vbk}. This is consistent with
our result.}
To look into this issue, one may use the rescaled variable
$\widetilde{\delta\chi}={\cal C}\delta\chi$ and expand the Lagrangian
to higher order in perturbations.
It would thus be interesting to explore the nonlinear dynamics of
the $\chi$ mode, which is however beyond the scope of this paper.

\section{Summary}

In this paper, we have proposed a scalar-tensor
reformulation of the recently proposed regularized version of
Lovelock gravity in four dimensions.
The effective scalar-tensor theory
is obtained by promoting the warp factor of the internal
$(D-4)$-dimensional space to a dynamical scalar field,
which survives after
performing the Kaluza-Klein reduction and
taking the $D\to 4$ limit
while keeping the rescaled coupling constants $\alpha_p'=(D-4)\alpha_p$
finite. The resultant theory resides in a
particular subclass of the Horndeski theory,
whose covariant field equations can be used for
any four-dimensional metric.
Our result is consistent with the conclusions of the different attempts
to derive the well-defined version of the four-dimensional
Einstein-Gauss-Bonnet theory~\cite{Fernandes:2020nbq,Hennigar:2020lsl,Bonifacio:2020vbk}.

Employing our scalar-tensor reformulation,
we have studied cosmological aspects of
regularized Einstein-Gauss-Bonnet gravity.
We have found that cosmological solutions generically
admit a scalar charge ${\cal C}$, which yields a radiation-like component.
All the previous cosmological results~\cite{Glavan:2019inb} are reproduced
at the level of the Lagrangian by taking ${\cal C}\to 0$.
However, in this limit the gravitational scalar degree of freedom
might be strongly coupled. This point needs further investigation.

\vspace{5mm}
{\bf Note added:}
While we were in the final stage of this work, the paper
by Lu and Pang~\cite{Lu:2020iav}
appeared in the arXiv, where the same idea of
reformulating regularized Einstein-Gauss-Bonnet gravity
is presented independently.
Their main focus is on black hole solutions,
while ours is on cosmological aspects.
Our conclusion agrees with them where they overlap.

\acknowledgments
We are grateful to Chunshan Lin for discussions.
The work of TK was supported by
MEXT KAKENHI Grant Nos.~JP16K17707, JP17H06359, and JP18H04355.


\bibliography{refs}
\bibliographystyle{JHEP.bst}
\end{document}